\documentclass[10pt]{bmc_article}

\usepackage{cite} 
\usepackage{url}  
\usepackage{ifthen}  
\usepackage{multicol}   
\usepackage[utf8]{inputenc} 
\usepackage{graphicx}
\usepackage{amsmath}
\usepackage{amssymb}
\usepackage{dsfont}
\usepackage{color}
\usepackage{epstopdf}

\newboolean{publ}

\newenvironment{bmcformat}{\baselineskip20pt\sloppy\setboolean{publ}{false}}{\baselineskip20pt\sloppy}

\begin{document}
\begin{bmcformat}


\title{The dynamics of alternative pathways to compensatory substitution}
 

\author{Chris A. Nasrallah\correspondingauthor$^1$%
         \email{Chris A. Nasrallah\correspondingauthor - nasrallah@statgen.ncsu.edu}
      }


\address{%
    \iid(1)Department of Genetics, North Carolina State University, Raleigh, NC 27695
}%

\maketitle


\begin{abstract}
The role of epistatic interactions among loci is a central question in evolutionary biology and is increasingly relevant in the genomic age.  While the population genetics of compensatory substitution have received considerable attention, most studies have focused on the case when natural selection is very strong against deleterious intermediates.  In the biologically-plausible scenario of weak to moderate selection there exist two alternate pathways for compensatory substitution. In one pathway, a deleterious mutation becomes fixed prior to occurrence of the compensatory mutation. In the other, the two loci are simultaneously polymorphic.  The rates of compensatory substitution along these two pathways and their relative probabilities are functions of the population size, selection strength, mutation rate, and recombination rate.  In this paper these rates and path probabilities are derived analytically and verified using population genetic simulations. The expected time durations of these two paths are similar when selection is moderate, but not when selection is weak.  The effect of recombination on the dynamics of the substitution process are explored using simulation.  Using the derived rates, a phylogenetic substitution model of the compensatory evolution process is presented that could be used for inference of population genetic parameters from interspecific data.  
\end{abstract}

\ifthenelse{\boolean{publ}}{\begin{multicols}{2}}{}



\section*{Background}

Complex interactions among loci play an important role in many evolutionary processes.  They have arisen in the context of adaptation as Sewall Wright's shifting balance theory \cite{wright31, wright32} and in speciation as Dobzhansky-Muller incompatibilities \cite{dobzhansky36, muller39}. Such interactions have been shown to play a role in the intramolecular evolution within proteins \cite{robinson03, rodrigue05} and {RNAs} \cite{chen99, yu06}, as well as the intermolecular co-evolution between different proteins \cite{clark09} and the regulation of gene expression \cite{goncalves12}.  An understanding of the the dynamics of such dependent evolutionary processes is therefore applicable for a wide range of fundamental questions in evolutionary biology.  

Often these problems are cast in the framework of a fitness landscape, with peaks corresponding to genotypes of high fitness and valleys to genotypes of low fitness \cite{huynen94, meer10}.  When the landscape is very flat and all genotypes of similar fitness, different loci may appear to evolve independently.  But when the landscape is more rugged and certain combinations of alleles are deleterious, the method and rate at which an entire population can cross from one peak to another become the central issues.  

If a haploid genome possesses a combination of alleles at different loci that are associated with high fitness, a mutation at any locus is likely to result in a suboptimal combination of alleles and consequently a haplotype of low fitness.  However, a second mutation at another locus may restore the combination of alleles to one of high fitness.  An entire population moving from one fitness peak to another in this manner is called compensatory substitution.

Theoretical population geneticists have used simple but powerful models, typically consisting of two loci with two alleles at each locus, to discern the rate at which such shifts can occur \cite{kimura85, iizuka96, michalakis96, phillips96, stephan96, higgs98, innan01, ichinose08, ichinose13}.  The expected time for these shifts is a function of the population size, the mutation rate, the strength of selection against intermediates, and the probability of recombination between loci \cite{kimura85}.  With only four allelic combinations, the frequency of these haplotypes in a population becomes a difficult three-dimensional diffusion problem \cite{stephan96, higgs98}.  The approach has uniformly been to simplify the dimensionality of the problem in order to obtain analytical results, with simulations frequently used to verify results and explore problems for which analytic solutions could not be found \cite{kimura85, iizuka96, michalakis96, phillips96, stephan96, higgs98, innan01, ichinose08, ichinose13}.

Kimura demonstrated that population genetic theory could allow for the occurrence of compensatory substitutions, using the simplest possible model of compensatory evolution with two linked loci and two alleles at each locus \cite{kimura85}.  Subsequent efforts have extended Kimura's results to models of greater complexity by allowing for:  different fitnesses for the two ``peak'' haplotypes \cite{iizuka96, michalakis96, iwasa04}, different fitnesses for the two deleterious intermediate haplotypes \cite{iizuka96, stephan96, michalakis96, innan01}, different mutation rates \cite{iizuka96, stephan96, innan01}, reversible mutations \cite{higgs98, innan01, ichinose13}, diploid populations \cite{ichinose08, ichinose13}, and expanding to a two locus four allele model \cite{higgs98}.  These mutation/selection models are depicted in Figure S1.

Each of these studies made simplifying assumptions about the strength of selection and mutation, and in order to compare their results it is useful to know in what ranges of parameter space their results may be applicable.  The figures from these papers and the simulations used by the authors are indicators of the range of parameter space in which these authors consider their approaches reasonable.  Figure 1 summarizes the approximate ranges of parameter space explored by each of these papers, based on their simulations and figures. This focuses on mutation and selection, both scaled by population size, although most of these papers explored recombination as well.

While most of this theoretical work assumed that selection (and often mutation) was very strong (but see \cite{innan01}), there are many cases in which selection against intermediates is not expected to be strong.  A high frequency of deleterious intermediate states was observed among homologous stem pairs of {RNA} between species \cite{rousset91}, taken as evidence that selection on {RNA} intermediates could not be as strong as previously supposed.  The role effective population size has played in the evolution of {tRNAs} was examined \cite{piskol11} and it was concluded that natural selection on paired sites has been weak to moderate.  It has also been observed in {RNA} intramolecular interactions that certain non-canonical intermediate configurations ({GU} or {UG}) are more stable than others, and that compensatory substitutions tend to favor paths through less-deleterious intermediates \cite{meer10}. These studies underscore the need to understand the dynamics of the compensatory substitution process when selection is moderate to weak. 

There are two pathways to compensatory substitution.  When selection is very weak there is a substantial chance (particularly if the population size is small) that a deleterious mutation can become fixed in the population before a second (compensatory) mutation arises and goes to fixation.  This will be referred to as a ``Type 1'' event (Fig.\ 2A).  However, when selection is very strong the deleterious haplotype cannot become fixed in the population.  In this case, compensatory substitution can only occur if the second mutation arises before the first mutation is lost due to selection, and the new high-fitness double mutant may then drift to fixation.  This kind of two-at-a-time substitution will be called a ``Type 2'' event (Fig.\ 2B).  When selection is of intermediate strength both of these substitution pathways are plausible (Fig.\ 2C).

The dynamics of the compensatory substitution process for both Type 1 and Type 2 events is the focus of this work.  The rates of these alternate events are derived, representing the first derivation of the direct double-substitution rate under weak selection and reversible mutation.  The interactions among selection strength, population size, mutation, recombination, and their effect on the probability of the alternative compensatory substitution pathways are explored using simulation.  This work also shows that the expected time to compensatory substitution may depend upon the fixation pathway.  These results yield insights into how fitness valleys can be crossed when selection is more moderate.

\section*{Methods} 

\medskip

\subsection*{Theory}

\medskip
\subsubsection*{Substitution Model}

Consider a population of $N$ diploid (or $2N$ haploid) individuals and a two-locus, two-allele mutation model as shown in Fig.\ 3A.  The first locus consists of alleles $A$ and $a$, the second of alleles $B$ and $b$.  Let $\mu$ be the bidirectional mutation rate between alleles $A$ and $a$, and let this same rate correspond to mutations between $B$ and $b$.  Under a haploid (genic) selection mechanism, the fitnesses of the four possible haplotypes $AB$, $aB$, $Ab$, and $ab$ are 1, $1-s$, $1-s$, and 1 respectively.  This situation corresponds to the case of symmetric neutral compensatory evolution \cite{kimura85} but with reversible mutation \cite{higgs98}.  For this derivation assume that the loci are tightly linked, such that recombination between loci does not occur. 

Define a four-state continuous-time Markov chain of the substitution process, as shown in Fig.\ 3B, in which each state represents the novel fixation of the population for each of these four haplotypes.  The notion of novel fixation events is an important one: a population fixed for $AB$, becoming polymorphic then losing the polymorphism to become again fixed for $AB$ does not represent a novel fixation event and no change of state will have been observed.
In this model the population may move between two fixed states if (1) the haplotypes differ at only one of the two loci, or (2) both haplotypes are of high fitness.  

Direct movement between states $Ab$ and $aB$ is disallowed, as such a move is highly unlikely unless intermediates are only very slightly deleterious.  This is because if both of these states are of low fitness, the process must actually avoid a fitness peak while making two substitutions.  If only single mutation events are allowed, then one of the fitter haplotypes must be created in the process, and is likely to fix via selection.  If double mutations are allowed, these will occur at rate $\mu^2$ and must then fix by drift, and it is far more likely that a single mutant of higher fitness will arise before this happens.

\subsubsection*{Rates into deleterious states}

Consider a population fixed for the $AB$ haplotype.  Following the formulation of \cite{halpern98}, the rate at which new $aB$ haplotypes arise that are destined to go to fixation in this population is the product of the rate at which new $aB$ mutants arise in the population and the probability that such a new mutation becomes fixed in the population.  The rate at which $aB$ haplotypes arise in a population of $2N$ $AB$ haplotypes, or equivalently the probability that a new $aB$ mutant will arise in such a population in a single generation, is $2N\mu$, where $\mu$ is the per-locus mutation rate.  Recalling that the $aB$ haplotype has fitness $1-s$, the probability of fixation of such a new mutation is given by $\frac{1-e^{2s}}{1-e^{4Ns}}$ \cite{kimura62}.  Consequently the rate from $AB$ to $aB$ is
	\begin{equation}\label{eq:r1}
		r_1 = 2 N \mu \frac{1-e^{2s}}{1-e^{4Ns}}.
	\end{equation}
If the assumption is made that mutation rates are uniform across loci and between all alleles, then Equation \ref{eq:r1} will hold for leaving any state with fitness $1$ \{$AB, ab$\} and entering any state with fitness $1-s$ \{$aB, Ab$\}.

The probability of fixation of a deleterious haplotype does not depend upon whether or not another fit haplotype arises in the process.  Consider a population of $AB$ individuals in which an $aB$ haplotype has arisen.  Before it becomes fixed, suppose a mutation occurs creating a new $ab$ haplotype from one of the copies of $aB$.  Because the $AB$ and $ab$ haplotypes are of equal fitness, the emergence of the new haplotype affects the probability of fixation of the intermediate exactly as would a reverse mutation.  Alternatively, the introduction of a new $Ab$ haplotype via mutation from $AB$ will serve to increase slightly the probability of fixation of the $aB$ haplotype by lowering the overall population fitness, but this effect is expected to be slight unless the mutation rate is very high.

\subsubsection*{Rates out of deleterious states}
 
Consider a population fixed for the $aB$ haplotype.  If the population fitness is renormalized to be $1$, then a new $ab$ haplotype arising in such a population will have fitness $1+t = \frac{1}{1-s}$, and will have selection coefficient  $t = \frac{s}{1-s}$ where $t > 0$ implies selection in favor of this haplotype.  In this case the probability of fixation of such a new haplotype given that it has arisen is $\frac{1-e^{-2t}}{1-e^{-4Nt}}$, and such haplotypes arise by mutation at rate $2N\mu$ per generation, yielding an overall rate of
	\begin{equation}\label{eq:r2}
		r_2 =  2 N \mu \frac{1-e^{-2t}}{1-e^{-4Nt}}.
	\end{equation}
As before, under symmetric mutation this rate $r_2$ will hold for leaving states with fitness $1-s$ \{$aB, Ab$\} and entering any state with fitness $1$ \{$AB, ab$\}.

\subsubsection*{Rates of double substitution}

For Equations \ref{eq:r1} and \ref{eq:r2} the assumption was made that new haplotypes arise in populations that are fixed for a particular state. It is possible however that a new haplotype could arise while the population is still polymorphic, and the identities and frequencies of the existing haplotypes could vary greatly.  This will be considered when deriving the rate of entering the $ab$ state.

With four possible haplotypes, obtaining formulae for the change in frequency of these haplotypes becomes a three-dimensional diffusion problem and is quite challenging.  However, in the case in which some haplotype fitnesses are equivalent the problem may be simplified.   By grouping states with equal fitness the dimensionality of the problem can be reduced \cite{innan01}.  Let $x$ be the combined frequency of the two deleterious haplotypes in the population (each with fitness $1-s$), and $y=1-x$ be the combined frequency of the remaining haplotypes (each with fitness $1$).  In this case, the problem reduces to a one-dimensional diffusion process on $x$.  Sewall Wright \cite{wright31} derived the equilibrium distribution of a mutant allele class resulting from the actions of reversible mutation and selection in a finite population as
	\begin{equation} \label{eq:Wright}
		\phi(x) = Ce^{4Nsx} x^{2\theta-1} (1-x)^{2\theta-1}
	\end{equation}
where $\theta = 4N\mu$ and $C$ is a normalizing constant such that $\int^1_0{\phi(x) dx = 1}$.  Equation \ref{eq:Wright} differs slightly from Wright's original because there are two loci and mutations may occur at either locus.  This provides the distribution of the frequency of deleterious intermediate haplotypes \{$Ab, aB$\}.

Consider a population lacking the $ab$ haplotype.  If the mutation rate is sufficiently low that the possibility of a double mutant $ab$ arising from an $AB$ haplotype in a single generation can be ignored, then $ab$ haplotypes can only arise from $Ab$ or $aB$ haplotypes via single mutation.  Therefore, conditional on the deleterious frequency $x$, the probability that a new $ab$ haplotype enters the population this generation is $2N\mu x$.

Under the assumption that the mutation rate is very small relative to the population size, the fixation probability of the $y$ group can be approximated by
	\begin{equation}\label{eq:p_y}
		p_y = \frac{ 1 - e^{-4Nsy'} } { 1 - e^{-4Ns} }
	\end{equation}
where $y'=y+\frac{1}{2N}$ is the frequency of the $y$ group after a new $ab$ mutant has entered the population \cite{kimura62}.  The probability of fixation of the new $ab$ haplotype is 
	\begin{equation}
		p_{ab} = \mathds{P}(ab \mbox{ fixes} \mid AB \mbox{ or } ab \mbox{ fixes}) 
			\mathds{P}(AB \mbox{ or } ab \mbox{ fixes}) 
			= \frac{p_y}{2Ny'} 
	\end{equation}
since $AB$ and $ab$ have equal fitness.  Taken together, and integrating over all possible values of the deleterious intermediate frequency, the average probability $\alpha$ in each generation that a new $ab$ haplotype will arise that is destined to go to fixation is
	\begin{align}
		\alpha &=  \int^1_0 { 
			\mathds{P}(ab \mbox{ arises} \mid x)  
			\mathds{P}(ab \mbox{ fixes} \mid x, ab \mbox{ arises}) 
			\mathds{P}(x) dx} \notag \\ 
		&= 2N \mu \int^1_0 { p_{ab} x\phi(x)dx} \label{eq:alpha}.
	\end{align}

Equations \ref{eq:p_y} - \ref{eq:alpha}, from \cite{innan01} but included here that this manuscript may be self-contained, give the rate $\alpha$ at which new successful $ab$ haplotypes enter the population.  By marginalizing over the stationary frequencies of deleterious intermediates at equilibrium, this key insight \cite{innan01} implicitly considers both pathways by which the new $ab$ haplotype can fix (with or without a deleterious intermediate ever fixing in the population).  Previous work on rates of compensatory substitution only considered the case in which the deleterious intermediate could never fix because selection was too strong \cite{kimura85, iizuka96, higgs98}.  

Considering the Markov substitution model of Fig.\ 3B, another way to describe Equation \ref{eq:alpha} is as the total rate \emph{into} state $ab$  
	\begin{align}
		\alpha &= \sum_{k \neq ab} { \pi_k r_{k \rightarrow ab}}\\
		&= (\pi_{aB} + \pi_{Ab}) r_2 + \pi_{AB} r_3 \notag
	\end{align}
where $\pi_k$ is the stationary probability of being in state $k$ and $r_2$ is given by Equation \ref{eq:r2}.  So the total rate into state $ab$ is the sum of rates into $ab$ from each other state, weighted by the stationary probability of being in that particular state.  This assumes that the detailed-balance conditions are met, meaning that the process is at stationarity.  This is not such a strong assumption, as the process reaches something very close to stationarity quickly \cite{innan01}.  Solving for $r_3$ yields
	\begin{equation}\label{eq:r3}
		r_3 = \frac{ \alpha - (\pi_{aB} + \pi_{Ab}) r_2 }{\pi_{AB}}.
	\end{equation}
Because the model is symmetric, $\alpha$ also describes the total rate into state $AB$, and therefore
	\begin{equation}\label{eq:r4}
		r_4 = \frac{ \alpha - (\pi_{aB} + \pi_{Ab}) r_2 }{\pi_{ab}}.
	\end{equation}

The expected stationary frequency of the fixed states can be derived from Equation \ref{eq:Wright}, which provides the distribution of the frequency of the pooled deleterious intermediate haplotypes (and conversely the pooled high-fitness haplotypes as well).  Alternatively these frequencies can be estimated from simulations.  Because the model is symmetric, $\pi_{AB} = \pi_{ab}$ and $\pi_{Ab} = \pi_{aB}$.

Equations \ref{eq:r3} and \ref{eq:r4} represent the rates of direct double-substitution, without first fixing the deleterious intermediate.  This is the first derivation of a direct double-substitution rate when selection is weak.

\subsubsection*{A discrete-time Markov model}

Embedded within the continuous-time Markov model of Figure 3B describing the rates of moving between states is a discrete-time Markov chain representing the transition probabilities between states (Fig.\ 3C).  Consider, as before, only novel fixation events such that each state is the population becoming newly fixed for that state.  The amount of time required for each fixation event to occur is not considered here, merely the state changes themselves.  This model will assist in understanding some of the properties of the compensatory substitution process.

For simplicity, the deleterious haplotypes $aB$ and $Ab$, both of relative fitness $1-s$, are considered together as a single state representing the population being fixed for one of the deleterious haplotypes.  Because both the $AB$ and $ab$ haplotypes have fitness 1 and because mutation is symmetric, the probabilities of moving from state $aB/Ab$ to either $AB$ or $ab$ are both $\frac{1}{2}$.  Since each simulation replicate ends when the population becomes fixed for the $ab$ haplotype, $ab$ is considered an absorbing state; a model of the true process differs only in that $ab$ is not absorbing and would be completely symmetric.

In this Markov model there is some probability $\beta$ of moving from state $AB$ directly to state $ab$ and some probability $1-\beta$ of moving to one of the deleterious intermediate states.  When $\beta=0$ the compensatory substitution can only occur via the fixation of the intermediate, and when $\beta=1$ the intermediate can never become fixed.

\subsection*{Simulations} 

The forward evolution of a population of $2N$ haploid individuals is simulated using Wright-Fisher sampling.  Each simulation replicate begins with the population fixed for the \emph{AB} haplotype and ends when the \emph{ab} haplotype becomes fixed in the population. The mutation-selection model used for simulation is the two-locus, two-allele model of \cite{higgs98}, shown in Figure 3A.  This is the simplest possible model of compensatory evolution that allows reversible mutation.   

In each generation the number of mutations occurring in the population in a generation is drawn as a Poisson($4N\mu$) random variable, where $\mu$ is the per generation per individual per locus mutation rate.  If more than one mutation is to be introduced, the haplotypes to be mutated are selected uniformly without replacement, meaning that two mutations cannot occur to the same copy.  Mutations to a haplotype occur at the first or second locus with equal probability.  The number of recombination events in each generation is drawn as a Poisson($2N\rho$) random variable, where $\rho$ is the recombination rate per generation per individual.  For each recombination event, two haplotypes are chosen without replacement with probability proportional to their frequency, the two selected haplotypes recombined, and the two recombinant haplotypes returned to the population.   Following mutation and recombination, one generation begets the next by resampling with replacement with probabilities proportional to the haplotype fitnesses (see Supporting Information for full details of population resampling).

After each round of mutation-recombination-resampling the composition of the population is evaluated  to determine the allele frequencies.  Whenever a haplotype becomes fixed in the population that is different from the last fixed haplotype, the generation and haplotype are recorded.  If the last fixed haplotype before the simulation is completed was $AB$, it is a Type 2 event (Fig.\ 2B).  If the last fixed haplotype before completion was a deleterious intermediate ($aB$ or $Ab$), it is a Type 1 event (Fig.\ 2A).

Type 2 events encompass two scenarios: either the deleterious intermediate haplotype is lost from the population before the $AB$ haplotype, leaving only the two high-fitness haplotypes, or the deleterious intermediate persists until after the $AB$ haplotype is lost.  While the latter case may result in one of the two new alleles becoming fixed before the other, as long as the intermediate haplotype does not fix it remains a Type 2 event (Fig.\ 2C).

Simulations were performed under combinations of a range of mutation rates, recombination rates, and selection coefficients (Table \ref{tab:WFsimParm}).  The simulation values chosen are similar to those of \cite{innan01}, and when scaled by the population size reflect reasonable estimates for a variety of natural populations.  One thousand simulation replicates were performed for each combination of parameters (see Supporting Information for exceptions).

\section*{Results and Discussion}

\medskip

\subsection*{Direct double substitution probability}

The direct double substitution probability $\beta$ (Eq.\ \ref{eq:beta}) can be easily estimated from simulations.  
When fixed for the $AB$ haplotype, the population can subsequently become fixed for either the $ab$ haplotype or for one of the deleterious intermediate haplotypes.  $\beta$ can be estimated as the proportion of times in which the population becomes fixed for the $ab$ haplotype after having been fixed for the $AB$ haplotype.

Estimates of $\beta$ under different simulation conditions are plotted as the points in Figure 4.  For small mutation rates and low selection coefficients there is very little chance of observing a direct double substitution, while for high mutation rates and large selection coefficients there is little chance of observing anything but direct double substitutions.

The expected value of $\beta$ can be derived from the rates of the continuous-time Markov model of Figure 3B.  Because the rates out of state $AB$ are given by Equations \ref{eq:r1} and \ref{eq:r3},
	\begin{equation}\label{eq:beta}
		\beta = \frac{ r_3 }{2r_{1} + r_3}.
	\end{equation}
	
Analytical expectations for $\beta$ from Equation \ref{eq:beta} are shown in Figure 4 for comparison to the simulation results.  The analytical result is a very good approximation to the observed direct double substitution probability across the range of selection coefficients, especially when the mutation rate is relatively low.  The poor approximation of the analytical prediction when the mutation rate is very high is to be expected since the derivation of $\alpha$ assumes relatively low mutation rates.

\subsection*{Alternate compensatory substitution pathway probabilities}

The probability that compensatory substitutions occur without the deleterious intermediate first becoming fixed (Type 2) is a function of selection, mutation, and recombination, all scaled by population size.  Importantly, it is not the same as the direct double-substitution probability $\beta$. This is because the process may visit other states before proceeding from $AB$ to $ab$ directly.  Consequently, $\beta$ serves as a lower bound on the probability that a compensatory substitution event is of Type 2.  

A more precise estimate for the probability of a compensatory substitution event being of Type 2 can be derived from $\beta$ as well.  Given that the population is in state $AB$, the probability that it moves directly to $ab$ without returning to $AB$ is $\beta$. The probability that it moves from $AB$ to $ab$ via the deleterious intermediate, without returning to $AB$ is $(1-\beta)/2$.  Thus the relative probability that it moves from $AB$ to $ab$ directly before moving to $ab$ via the deleterious intermediate is 
	\begin{equation} \label{eq:propType2}
	 \mathds{P}(\text{Type 2}) = \frac{\beta}{\beta+(1-\beta)/2} = \frac{2\beta}{1+\beta}.
	\end{equation}  
The population could of course have returned to the initial state $AB$ from the deleterious intermediate as well.  But because the process is Markovian, the probability of observing a Type 2 event given that the current state is $AB$ is still given by Equation \ref{eq:propType2}.

The analytical prediction of Equation \ref{eq:propType2} fits the results of simulations in the absence of recombination quite well across the range of selection coefficients explored, except (unsurprisingly) when the mutation rate is very high.  Like the direct double substitution probability $\beta$, the proportion of Type 2 compensatory substitution events increases with the strength of selection (Fig.\ 5).  In addition, Type 2 compensatory substitution events are very rare when the mutation rate is relatively low, such that the vast majority of compensatory substitution events involve the fixing of the deleterious intermediate haplotype. 

Recombination is often thought of as having the positive effect of putting together beneficial mutations \cite{fisher30, muller32}. Even negatively-epistatic alleles that are beneficial when paired will be aided by recombination \cite{michalakis96}. But when the two mutants are individually deleterious and jointly neutral, recombination increases the time required for compensatory evolution \cite{kimura85}.  Empirically, fewer compensatory substitutions are observed as physical distance (and hence recombination) between interacting loci increases \cite{stephan93, piskol08}.  

In this context, recombination reduces the proportion of Type 2 substitutions observed (Fig.\ 5).  Since the $a$ and $b$ alleles spend most of their time at low frequency due to selection against them, recombination serves to break apart $ab$ haplotypes much more than it creates them.  Consider a population with a single copy of each of the $a$ and $b$ alleles.  A single $ab$ haplotype will be broken up by recombination if it is one of the two selected, but an $ab$ haplotype will only be created by recombination if \emph{both} of the haplotypes bearing the deleterious alleles are selected. This tendency to separate rare alleles increases the chance that at least one of the alleles will be lost, decreasing the probability of Type 2 events. This effect of recombination is obviated when the mutation rate is very high, where Type 2 events dominate over a wide range of selective strength.

Taken together these paint a picture of the dynamics of the compensatory neutral substitution process.  If mutation rates are relatively low, then times to compensatory substitution will be very large, particularly when selection is strong \cite{stephan96}.  In this case the process will likely proceed through the deleterious intermediate fixed state, unless selection is strong (Fig.\ 5).  Recombination, if strong, will increase both the time to fixation \cite{innan01} and the probability of passing through the deleterious intermediate state (Fig.\ 5).  Conversely, high mutation and strong selection will lead to faster Type 2 substitution events \cite{kimura85}, and high mutation rate alone provides immunity from the effect of recombination (Fig.\ 5).  This suggests potentially very different compensatory substitution dynamics in different species.

\subsection*{Reversions to the ancestral state}

It was noted earlier that the population, after fixing a deleterious intermediate state, could revert to the original fixed state $AB$ rather than proceeding to the alternative state $ab$.  In fact, it can cycle back a number of times before eventually becoming fixed for $ab$, either via the intermediate or directly.  Each time in state $AB$ the probability of ``failure'' (leaving then returning to $AB$) is $(1-\beta)/2$, and the probability of ``success'' (reaching $ab$ by any means) is $(1+\beta)/2$.  The number of returns to $AB$ before ultimately reaching $ab$ is therefore a geometrically distributed random variable on the space \{0,1,2,...\} with mean $(1-\beta)/(1+\beta)$.  Because only the final path from $AB$ to $ab$ determines if an event is of Type 1 or Type 2, the expected number of reversions is the same for both pathways.

Figure 6 shows the average number of times in simulation that the population fixes a deleterious intermediate state and then returns to state $AB$ before fixing in state $ab$ via any pathway.  The number of returns to the initial state is lower when the mutation is high.  
This is because high mutation increases the rate of appearance (and subsequent fixation) of $ab$ haplotypes (Fig.\ 5); if intermediates do not become fixed, there can be no return to the initial condition. 
Recombination had little effect on the number of reversions, slightly increasing the number of reversions when selection is moderate to large and mutation is low (Fig.\ S2).  Increasing selection against intermediates reduces the number of reversions by removing deleterious alleles from the population; by preventing deleterious mutants from fixing in the population it also prevents reversions (as well as Type 1 fixation events).

\subsection*{Relative time of the two pathways}

An important question is whether the expected amount of time to compensatory substitution differs for Type 1 and Type 2 events.  It is not obvious that these two events should take the same amount of time, or if differences between them should remain constant across a wide range of selection, mutation, and recombination.  Because the expected number of reversions to the ancestral state is the same regardless of the pathway by which fixation is ultimately achieved, the power to detect differences between Type 1 and Type 2 events is maximized by comparing the time required for the final path from $AB$ to $ab$.  

The number of generations from the last time the population enters state $AB$ until the time it enters state $ab$ was calculated for each simulation replicate.  If there is genuinely no difference in the expected time for Type 1 and Type 2 fixation events, then the difference in the two sample means is expected to be zero.  Under certain assumptions about the samples (normally distributed and equal variances), the difference in means, normalized by the pooled variance, will be a standard normal random variable.  Large deviations in this test statistic imply significant differences in the expected times of the two pathways. 

The important observation is that natural selection against deleterious intermediates eliminates the difference in time required between the two paths observed, and this appears true across all mutation rates examined (Fig.\ 7).  With increased selection the probability of Type 2 events increases (Fig.\ 5), but although Type 1 events are less likely to occur under strong selection, when they do occur they take the same amount of time as Type 2 events.  This implies that the time to fixation alone is insufficient for determining the pathway by which fixation occurred if selection has been moderate. However, if selection has been very weak and the mutation rate is very high, it might be possible to use time to fixation to determine the fixation pathway.

Recombination had little effect on either the differences observed under weak selection or the lack of differences observed at stronger selection (Fig.\ 7).  Because recombination most strongly affects the probability of observing Type 2 events when the mutation rate is low and selection is strong (Fig.\ 5), if recombination is to have an effect on the difference in path times it should be most evident under these conditions.  However, it is under precisely these conditions that the difference between path times is minimized (Fig.\ 7).

\subsection*{Extensions}

\medskip

\subsubsection*{Model asymmetry}

The total rate of compensatory substitution $\alpha$ in the more general case in which mutation for the two loci and selection coefficients against the two deleterious haplotypes could differ was previously derived \cite{innan01}.  This is attractive because it would allow for more biologically realistic scenarios, such as the semi-stable U-G intermediates in {RNA} stem pairing \cite{rousset91}.  However, in deriving the more general case they made the assumption that $Ns$ was very large in order to rule out the possibility of the coexistence of both deleterious haplotypes.  This assumption might be problematic for many systems in which either population sizes or selection coefficients are not expected to be very large.  Nevertheless, it should be straightforward to incorporate into the work presented here.

The case in which the fitness of $ab$ is far greater than that of any other state, such as when normal cells mutate to cancerous ones, was considered by \cite{iwasa04}.  Under these conditions they derived a rate of double substitution, giving it the colorful term ``stochastic tunneling.''  However, unlike the approach shown here, these authors assumed that the probability of fixation of a new double mutant was independent of the population composition at the time it arises.  This assumption simplifies matters greatly, but such an approximation may be limited to the case in which the $ab$ state is of exceptionally high fitness.

\subsubsection*{Expanding the state space}

The two-locus, two-allele model considered here will be insufficient for some important biological cases, such as the evolution of {RNA} stem positions. In this case, either the state space of the model can be expanded to a two-locus, four-allele model, or the state space of {RNA} can be compressed somehow to fit the current model.  The latter approach	 is simpler, but risks losing valuable information about the evolutionary process by grouping states together.  The former approach for the case of strong selection and large populations was explored by \cite{higgs98}.

\subsubsection*{A Phylogenetic substitution model}

Because the rates derived in this work are rates of the substitution process, a phylogenetic model may be defined using these rates. In this case, the elements of the instantaneous rate matrix $\mathbf{Q} = \{ q_{ij} \}$ of the continuous-time Markov chain can be defined as
	\begin{equation} \label{eq:Q}
		q_{ij} = 
		\begin{cases} 
			u \pi_j r_1 			&\text{if $i \in \mathbb{Y}$ and $j \in \mathbb{X}$ }\\ 
			u \pi_j r_2 			&\text{if $i \in \mathbb{X}$ and $j \in \mathbb{Y}$ }\\ 
			u \pi_j r_3 			&\text{if $i = AB$ and $j = ab$ }\\ 
			u \pi_j r_4 			&\text{if $i = ab$ and $j = AB$ }\\ 
			0				&\text{if $i, j \in \mathbb{X}$ and $i \neq j$}\\ 
			-\displaystyle\sum_{i \neq j} q_{ij}	&\text{if $i=j$} 
		\end{cases}	
	\end{equation}
or alternatively, in terms of the population genetic parameters as		
		\begin{equation} \label{eq:Q2}
		q_{ij} = 
		\begin{cases} 
			u \pi_j 2 N \mu \frac{1-e^{2s}}{1-e^{4Ns}} 			
				&\text{if $i \in \mathbb{Y}$ and $j \in \mathbb{X}$ }\\ 
			u \pi_j 2 N \mu \frac{1-e^{-2s}}{1-e^{-4Ns}} 
					&\text{if $i \in \mathbb{X}$ and $j \in \mathbb{Y}$ }\\ 
			u \frac{\pi_{j}}{\pi_{i}} 2N\mu
			   \left(  
				\int_0^1  { 
				    \frac{x (1-e^{-2s(2N(1-x)+1))}}{(1-e^{-4Ns})(2N(1-x)+1)} 
				    \frac{e^{4Nsx} (x(1-x))^{8N\mu-1}}{\int_0^1 e^{4Nsz} (z(1-z))^{8N\mu-1} dz} dx 
				}   
			   	-  \frac{(1-e^{-2s}) \pi_{\mathbb{X}}} {1-e^{-4Ns}} 
			   \right)	
			   	&\text{if $i, j \in \mathbb{Y}$ and $i \neq j$}\\
			0				&\text{if $i, j \in \mathbb{X}$ and $i \neq j$}\\ 
			-\displaystyle\sum_{i \neq j} q_{ij}	&\text{if $i=j$} 
		\end{cases} 
	\end{equation} 
where $\mathbb{X} = \{aB, Ab\}$, $\mathbb{Y} = \{AB, ab\}$, and $u$ is a rate scaling factor to ensure that branch lengths are in terms of expected number of substitutions per site.  

Phylogenetic models of the compensatory substitution process have been developed previously \cite{tillier94, tillier95}, but these models have double and single substitution rates that are uncoupled from each other and that are purely descriptive in nature.  Equation \ref{eq:Q2} on the other hand represents a fully-defined phylogenetic substitution model for the analysis of compensatory substitution specified entirely in terms of the underlying population genetic parameters.  This is appealing because it moves towards models that are \emph{interpretive}, in which parameters have meaning with respect to the process of evolution, rather than models that are simply \emph{descriptive} of patterns observed.  

The model described by Equation \ref{eq:Q2} presents a potentially powerful way to investigate the evolution of compensatory evolution using interspecific data, allowing the estimation of population parameters of interest.  Application of the basic model described here to interspecific data would be making key assumptions about the evolutionary process, namely that each branch of the phylogenetic tree would share a common population size, selection coefficient, and mutation rate.  Relaxing this constraint to allow different branches to have their own parameters has been done in different contexts using the Dirichlet process \cite{huelsenbeck06, heath12} and could readily be applied to this model as well.

\subsubsection*{Origination vs. fixation process}

The total compensatory substitution rate $\alpha$ is actually the rate at which successful compensatory haplotypes \emph{arise} in the population, the inverse of which being the time until such haplotypes arise \cite{innan01}. This is slightly different from what was calculated in simulations, which was the time until the actual fixation event itself.  The difference in these times is very small: A compensatory (second) mutation that is destined to go to fixation, from the time it arises in the population, takes on the order of $4N$ generations until it becomes fixed \cite{kimura70, innan01}, which is trivial compared to the time typically spent waiting for the eventually-successful mutation to arise.  Consequently the simulations should be expected to yield ever-so-slightly longer fixation times than predicted by $\alpha$.  

These two processes are distinguished as origination and fixation processes, respectively \cite{gillespie91}. Origination processes can often have more desirable statistical properties than fixation processes. For example, in the case of neutral linkage the origination process remains Poisson while the fixation process is something more complex and overdispersed \cite{watterson82}.  While substitution models are rarely explicit about the process being modeled, the theoretical results and phylogenetic substitution model presented here pertain to the origination process rather than the fixation process.

\section*{Conclusions}

There are many cases in which loci interact epistatically and natural selection is weak to moderate, and this work derived rates for the compensatory neutral substitution process along the different possible paths to fixation under these conditions. These were shown to have good predictive power regarding the probability of observing compensatory substitution via alternate paths under a range of conditions.  Because there is little difference in the time to fixation between the paths when selection is not negligible, the relative probability of these paths may be important for distinguishing the evolutionary history of compensatory substitutions.  To this end, a phylogenetic substitution model parameterized in terms of these population genetic parameters was presented that could be used for inference of population histories using interspecific data.


\section*{Acknowledgements}
  \ifthenelse{\boolean{publ}}{\small}{}
The author would like to thank Jeff Thorne, Josh Schraiber, Alex Griffing and two anonymous reviewers for helpful comments on this manuscript.  This work was supported by National Institutes of Health grants GM-069801 (to John Huelsenbeck) and GM-070806 (to Jeff Thorne), and National Science Foundation postdoctoral fellowship DBI-1202884 (to CN). 


\newpage
{\ifthenelse{\boolean{publ}}{\footnotesize}{\small}
 \bibliographystyle{bmc_article}  
 \bibliography{bayes} }     


\begin{thebibliography}{10}
\providecommand{\url}[1]{[#1]}
\providecommand{\urlprefix}{}

\bibitem{wright31}
Wright S: \textbf{Evolution in {M}endelian populations}. \emph{Genetics} 1931,
  \textbf{16}:97--159.

\bibitem{wright32}
Wright S: \textbf{The roles of mutation, inbreeding, crossbreeding and
  selection in evolution}. \emph{Proceedings of the Sixth International
  Congress on Genetics} 1932, \textbf{1}:356--366.

\bibitem{dobzhansky36}
Dobzhansky T: \textbf{Studies on hybrid sterility. {II}. {L}ocalization of
  sterility factors in \textit{{D}rosophila} pseudoobscura hybrids}.
  \emph{Genetics} 1936, \textbf{21}:113--135.

\bibitem{muller39}
Muller HJ: \textbf{Reversibility in evolution considered from the standpoint of
  genetics}. \emph{Bio. Rev. Camb. Philos. Soc.} 1939, \textbf{14}:261--280.

\bibitem{robinson03}
Robinson DM, Jones DT, Kishino H, Goldman N, Thorne JL: \textbf{Protein
  evolution with dependence among codons due to tertiary structure}.
  \emph{Molecular Biology and Evolution} 2003, \textbf{20}:1692--1704.

\bibitem{rodrigue05}
Rodrigue N, Lartillot N, Bryant D, Philippe H: \textbf{Site interdependence
  attributed to tertiary structure in amino acid sequence evolution}.
  \emph{Gene} 2005, \textbf{347}:207--217.

\bibitem{chen99}
Chen Y, Carlini DB, Baines JF, Parsch J, Braverman JM, Tanda S, Stephan W:
  \textbf{{RNA} secondary structure and compensatory evolution}. \emph{Genes
  Genet. Syst.} 1999, \textbf{74}:271--286.

\bibitem{yu06}
Yu J, Thorne JL: \textbf{Dependence among sites in {RNA} evolution}.
  \emph{Molecular Biology and Evolution} 2006, \textbf{23}:1525--1537.

\bibitem{clark09}
Clark NL, Gasper J, Sekino M, Springer SA, Aquadro CF, Swanson WJ:
  \textbf{Coevolution of interacting fertilization proteins}. \emph{PLoS
  Genetics} 2009, \textbf{5}:e1000570.

\bibitem{goncalves12}
Goncalves A, Leigh-Brown S, Thybert D, Stefflova K, Turro E, Flicek P, Brazma
  A, Odom DT, Marioni JC: \textbf{Extensive compensatory cis-trans regulation
  in the evolution of mouse gene expression}. \emph{Genome Research} 2012,
  \textbf{22}:2376--2384.

\bibitem{huynen94}
Huynen M, Hogeweg P: \textbf{Pattern generation in molecular evolution:
  exploitation of the variation in {RNA} landscapes}. \emph{Journal of
  Molecular Evolution} 1994, \textbf{39}:71--79.

\bibitem{meer10}
Meer MV, Kondrashov AS, Artzy-Randrup Y, Kondrashov FA: \textbf{Compensatory
  evolution in mitochondrial t{RNA}s navigates valleys of low fitness}.
  \emph{Nature} 2010, \textbf{464}:279--282.

\bibitem{kimura85}
Kimura M: \textbf{The role of compensatory neutral mutations in molecular
  evolution}. \emph{Journal of Genetics} 1985, \textbf{64}:7--19.

\bibitem{iizuka96}
Iizuka M, Takefu M: \textbf{Average time until fixation of mutants with
  compensatory fitness interaction}. \emph{Genes Genet. Syst.} 1996,
  \textbf{71}:167--173.

\bibitem{michalakis96}
Michalakis Y, Slatkin M: \textbf{Interaction of selection and recombination in
  the fixation of negative-epistatic genes}. \emph{Genet. Res.} 1996,
  \textbf{67}:257--269.

\bibitem{phillips96}
Phillips PC: \textbf{Waiting for a compensatory mutation: phase zero of the
  shifting-balance process}. \emph{Genetical Research} 1996,
  \textbf{67}:271--283.

\bibitem{stephan96}
Stephan W: \textbf{The rate of compensatory evolution}. \emph{Genetics} 1996,
  \textbf{144}:419--426.

\bibitem{higgs98}
Higgs PG: \textbf{Compensatory neutral mutations and the evolution of {RNA}}.
  \emph{Genetica} 1998, \textbf{102/103}:91--101.

\bibitem{innan01}
Innan H, Stephan W: \textbf{Selection intensity against deleterious mutations
  in {RNA} secondary structures and rate of compensatory nucleotide
  substitutions}. \emph{Genetics} 2001, \textbf{159}:389--399.

\bibitem{ichinose08}
Ichinose M, Iizuka M, Kado T, Takefu M: \textbf{Compensatory evolution in
  diploid populations}. \emph{Theoretical Population Biology.} 2008,
  \textbf{74}:199--207.

\bibitem{ichinose13}
Ichinose M, Iizuka M, Kusumi J, Takefu M: \textbf{Models of compensatory
  molecular evolution: effects of back mutation}. \emph{Theoretical Population
  Biology.} 2013, \textbf{323}:1--10.

\bibitem{iwasa04}
Iwasa Y, Michor F, Nowak MA: \textbf{Stochastic tunnels in evolutionary
  dynamics}. \emph{Genetics} 2004, \textbf{166}:1571--1579.

\bibitem{rousset91}
Rousset F, Pelandakis M, Solignac M: \textbf{Evolution of compensatory
  substitutions through G-U intermediate state in \textit{{D}rosophila}
  r{RNA}}. \emph{PNAS} 1991, \textbf{88}:10032--10036.

\bibitem{piskol11}
Piskol R, Stephan W: \textbf{The role of the effective population size in
  compensatory evolution}. \emph{Genome Biology and Evolution} 2011,
  \textbf{3}:528--538.

\bibitem{halpern98}
Halpern AL, Bruno WJ: \textbf{Evolutionary distances for protein-coding
  sequences: {M}odeling site-specific residue frequencies}. \emph{Molecular
  Biology and Evolution} 1998, \textbf{15}:910--917.

\bibitem{kimura62}
Kimura M: \textbf{On the probability of fixation of mutant genes in a
  population}. \emph{Genetics} 1962, \textbf{47}:713--719.

\bibitem{fisher30}
Fisher RA: \emph{The {G}enetical {T}heory of {N}atural {S}election}. Oxford:
  Oxford University Press 1930.

\bibitem{muller32}
Muller HJ: \textbf{Some genetic aspects of sex}. \emph{The American Naturalist}
  1932, \textbf{66}:118--138.

\bibitem{stephan93}
Stephan W, Kirby DA: \textbf{{RNA} folding in \textit{{D}rosophila} shows a
  distance effect for compensatory fitness interactions}. \emph{Genetics} 1993,
  \textbf{135}:97--103.

\bibitem{piskol08}
Piskol R, Stephan W: \textbf{Analyzing the evolution of {RNA} secondary
  structures in vertebrate introns using {K}imura's model of compensatory
  fitness interactions}. \emph{Molecular Biology and Evolution} 2008,
  \textbf{25}:2483--2492.

\bibitem{tillier94}
Tillier ERM: \textbf{Maximum likelihood with multiparameter models of
  substitution}. \emph{Journal of Molecular Evolution} 1994,
  \textbf{39}:409--417.

\bibitem{tillier95}
Tillier ERM, Collins RA: \textbf{Neighbor joining and maximum likelihood with
  RNA sequences: addressing the interdependence of sites}. \emph{Molecular
  Biology and Evolution} 1995, \textbf{12}:7--15.

\bibitem{huelsenbeck06}
Huelsenbeck JP, Jain S, Frost SWD, Pond SLK: \textbf{A {D}irichlet process
  model for detecting positive selection in protein-coding {DNA} sequences}.
  \emph{Proceedings of the National Academy of Science, U.S.A.} 2006,
  \textbf{103}:6263--6268.

\bibitem{heath12}
Heath TA, Holder MT, Huelsenbeck JP: \textbf{A {D}irichlet process prior for
  estimating lineage-specific substitution rates}. \emph{Molecular Biology and
  Evolution} 2012, \textbf{29}:939--955.

\bibitem{kimura70}
Kimura M: \textbf{The length of time required for a selectively neutral mutant
  to reach fixation through random frequency drift in a finite population}.
  \emph{Genet. Res.} 1970, \textbf{15}:131--133.

\bibitem{gillespie91}
Gillespie JH: \emph{The causes of molecular evolution}. New York: Oxford
  University Press 1991.

\bibitem{watterson82}
Watterson GA: \textbf{Substitution times for mutant nucleotides}. \emph{Journal
  of Applied Probability} 1982, \textbf{19}:59--70.

\end{thebibliography}

\newcommand{\BMCxmlcomment}[1]{}

\BMCxmlcomment{

<refgrp>

<bibl id="B1">
  <title><p>Evolution in {M}endelian populations</p></title>
  <aug>
    <au><snm>Wright</snm><fnm>S.</fnm></au>
  </aug>
  <source>Genetics</source>
  <pubdate>1931</pubdate>
  <volume>16</volume>
  <fpage>97</fpage>
  <lpage>-159</lpage>
</bibl>

<bibl id="B2">
  <title><p>The roles of mutation, inbreeding, crossbreeding and selection in
  evolution</p></title>
  <aug>
    <au><snm>Wright</snm><fnm>S.</fnm></au>
  </aug>
  <source>Proceedings of the Sixth International Congress on Genetics</source>
  <pubdate>1932</pubdate>
  <volume>1</volume>
  <fpage>356</fpage>
  <lpage>-366</lpage>
</bibl>

<bibl id="B3">
  <title><p>Studies on hybrid sterility. {II}. {L}ocalization of sterility
  factors in \textit{{D}rosophila} pseudoobscura hybrids</p></title>
  <aug>
    <au><snm>Dobzhansky</snm><fnm>T.</fnm></au>
  </aug>
  <source>Genetics</source>
  <pubdate>1936</pubdate>
  <volume>21</volume>
  <fpage>113</fpage>
  <lpage>-135</lpage>
</bibl>

<bibl id="B4">
  <title><p>Reversibility in evolution considered from the standpoint of
  genetics</p></title>
  <aug>
    <au><snm>Muller</snm><fnm>H. J.</fnm></au>
  </aug>
  <source>Bio. Rev. Camb. Philos. Soc.</source>
  <pubdate>1939</pubdate>
  <volume>14</volume>
  <fpage>261</fpage>
  <lpage>-280</lpage>
</bibl>

<bibl id="B5">
  <title><p>Protein evolution with dependence among codons due to tertiary
  structure</p></title>
  <aug>
    <au><snm>Robinson</snm><fnm>D. M.</fnm></au>
    <au><snm>Jones</snm><fnm>D. T.</fnm></au>
    <au><snm>Kishino</snm><fnm>H.</fnm></au>
    <au><snm>Goldman</snm><fnm>N.</fnm></au>
    <au><snm>Thorne</snm><fnm>J. L.</fnm></au>
  </aug>
  <source>Molecular Biology and Evolution</source>
  <pubdate>2003</pubdate>
  <volume>20</volume>
  <fpage>1692</fpage>
  <lpage>-1704</lpage>
</bibl>

<bibl id="B6">
  <title><p>Site interdependence attributed to tertiary structure in amino acid
  sequence evolution</p></title>
  <aug>
    <au><snm>Rodrigue</snm><fnm>N.</fnm></au>
    <au><snm>Lartillot</snm><fnm>N.</fnm></au>
    <au><snm>Bryant</snm><fnm>D.</fnm></au>
    <au><snm>Philippe</snm><fnm>H.</fnm></au>
  </aug>
  <source>Gene</source>
  <pubdate>2005</pubdate>
  <volume>347</volume>
  <fpage>207</fpage>
  <lpage>-217</lpage>
</bibl>

<bibl id="B7">
  <title><p>{RNA} secondary structure and compensatory evolution</p></title>
  <aug>
    <au><snm>Chen</snm><fnm>Y.</fnm></au>
    <au><snm>Carlini</snm><fnm>D. B.</fnm></au>
    <au><snm>Baines</snm><fnm>J. F.</fnm></au>
    <au><snm>Parsch</snm><fnm>J.</fnm></au>
    <au><snm>Braverman</snm><fnm>J. M.</fnm></au>
    <au><snm>Tanda</snm><fnm>S.</fnm></au>
    <au><snm>Stephan</snm><fnm>W.</fnm></au>
  </aug>
  <source>Genes Genet. Syst.</source>
  <pubdate>1999</pubdate>
  <volume>74</volume>
  <fpage>271</fpage>
  <lpage>-286</lpage>
</bibl>

<bibl id="B8">
  <title><p>Dependence among sites in {RNA} evolution</p></title>
  <aug>
    <au><snm>Yu</snm><fnm>J.</fnm></au>
    <au><snm>Thorne</snm><fnm>J. L.</fnm></au>
  </aug>
  <source>Molecular Biology and Evolution</source>
  <pubdate>2006</pubdate>
  <volume>23</volume>
  <fpage>1525</fpage>
  <lpage>-1537</lpage>
</bibl>

<bibl id="B9">
  <title><p>Coevolution of interacting fertilization proteins</p></title>
  <aug>
    <au><snm>Clark</snm><fnm>N. L.</fnm></au>
    <au><snm>Gasper</snm><fnm>J.</fnm></au>
    <au><snm>Sekino</snm><fnm>M.</fnm></au>
    <au><snm>Springer</snm><fnm>S. A.</fnm></au>
    <au><snm>Aquadro</snm><fnm>C. F.</fnm></au>
    <au><snm>Swanson</snm><fnm>W. J.</fnm></au>
  </aug>
  <source>PLoS Genetics</source>
  <pubdate>2009</pubdate>
  <volume>5</volume>
  <fpage>e1000570</fpage>
</bibl>

<bibl id="B10">
  <title><p>Extensive compensatory cis-trans regulation in the evolution of
  mouse gene expression</p></title>
  <aug>
    <au><snm>Goncalves</snm><fnm>A.</fnm></au>
    <au><snm>Leigh Brown</snm><fnm>S.</fnm></au>
    <au><snm>Thybert</snm><fnm>D.</fnm></au>
    <au><snm>Stefflova</snm><fnm>K.</fnm></au>
    <au><snm>Turro</snm><fnm>E.</fnm></au>
    <au><snm>Flicek</snm><fnm>P.</fnm></au>
    <au><snm>Brazma</snm><fnm>A.</fnm></au>
    <au><snm>Odom</snm><fnm>D. T.</fnm></au>
    <au><snm>Marioni</snm><fnm>J. C.</fnm></au>
  </aug>
  <source>Genome Research</source>
  <pubdate>2012</pubdate>
  <volume>22</volume>
  <fpage>2376</fpage>
  <lpage>-2384</lpage>
</bibl>

<bibl id="B11">
  <title><p>Pattern generation in molecular evolution: exploitation of the
  variation in {RNA} landscapes</p></title>
  <aug>
    <au><snm>Huynen</snm><fnm>M.A.</fnm></au>
    <au><snm>Hogeweg</snm><fnm>P.</fnm></au>
  </aug>
  <source>Journal of Molecular Evolution</source>
  <pubdate>1994</pubdate>
  <volume>39</volume>
  <fpage>71</fpage>
  <lpage>-79</lpage>
</bibl>

<bibl id="B12">
  <title><p>Compensatory evolution in mitochondrial t{RNA}s navigates valleys
  of low fitness</p></title>
  <aug>
    <au><snm>Meer</snm><fnm>M. V.</fnm></au>
    <au><snm>Kondrashov</snm><fnm>A. S.</fnm></au>
    <au><snm>Artzy Randrup</snm><fnm>Y.</fnm></au>
    <au><snm>Kondrashov</snm><fnm>F. A.</fnm></au>
  </aug>
  <source>Nature</source>
  <pubdate>2010</pubdate>
  <volume>464</volume>
  <fpage>279</fpage>
  <lpage>-282</lpage>
</bibl>

<bibl id="B13">
  <title><p>The role of compensatory neutral mutations in molecular
  evolution</p></title>
  <aug>
    <au><snm>Kimura</snm><fnm>M.</fnm></au>
  </aug>
  <source>Journal of Genetics</source>
  <pubdate>1985</pubdate>
  <volume>64</volume>
  <fpage>7</fpage>
  <lpage>-19</lpage>
</bibl>

<bibl id="B14">
  <title><p>Average time until fixation of mutants with compensatory fitness
  interaction</p></title>
  <aug>
    <au><snm>Iizuka</snm><fnm>M.</fnm></au>
    <au><snm>Takefu</snm><fnm>M.</fnm></au>
  </aug>
  <source>Genes Genet. Syst.</source>
  <pubdate>1996</pubdate>
  <volume>71</volume>
  <fpage>167</fpage>
  <lpage>-173</lpage>
</bibl>

<bibl id="B15">
  <title><p>Interaction of selection and recombination in the fixation of
  negative-epistatic genes</p></title>
  <aug>
    <au><snm>Michalakis</snm><fnm>Y.</fnm></au>
    <au><snm>Slatkin</snm><fnm>M.</fnm></au>
  </aug>
  <source>Genet. Res.</source>
  <pubdate>1996</pubdate>
  <volume>67</volume>
  <fpage>257</fpage>
  <lpage>-269</lpage>
</bibl>

<bibl id="B16">
  <title><p>Waiting for a compensatory mutation: phase zero of the
  shifting-balance process</p></title>
  <aug>
    <au><snm>Phillips</snm><fnm>P. C.</fnm></au>
  </aug>
  <source>Genetical Research</source>
  <pubdate>1996</pubdate>
  <volume>67</volume>
  <fpage>271</fpage>
  <lpage>-283</lpage>
</bibl>

<bibl id="B17">
  <title><p>The rate of compensatory evolution</p></title>
  <aug>
    <au><snm>Stephan</snm><fnm>W.</fnm></au>
  </aug>
  <source>Genetics</source>
  <pubdate>1996</pubdate>
  <volume>144</volume>
  <fpage>419</fpage>
  <lpage>-426</lpage>
</bibl>

<bibl id="B18">
  <title><p>Compensatory neutral mutations and the evolution of
  {RNA}</p></title>
  <aug>
    <au><snm>Higgs</snm><fnm>P. G.</fnm></au>
  </aug>
  <source>Genetica</source>
  <pubdate>1998</pubdate>
  <volume>102/103</volume>
  <fpage>91</fpage>
  <lpage>-101</lpage>
</bibl>

<bibl id="B19">
  <title><p>Selection intensity against deleterious mutations in {RNA}
  secondary structures and rate of compensatory nucleotide
  substitutions</p></title>
  <aug>
    <au><snm>Innan</snm><fnm>H.</fnm></au>
    <au><snm>Stephan</snm><fnm>W.</fnm></au>
  </aug>
  <source>Genetics</source>
  <pubdate>2001</pubdate>
  <volume>159</volume>
  <fpage>389</fpage>
  <lpage>-399</lpage>
</bibl>

<bibl id="B20">
  <title><p>Compensatory evolution in diploid populations</p></title>
  <aug>
    <au><snm>Ichinose</snm><fnm>M.</fnm></au>
    <au><snm>Iizuka</snm><fnm>M.</fnm></au>
    <au><snm>Kado</snm><fnm>T.</fnm></au>
    <au><snm>Takefu</snm><fnm>M.</fnm></au>
  </aug>
  <source>Theoretical Population Biology.</source>
  <pubdate>2008</pubdate>
  <volume>74</volume>
  <fpage>199</fpage>
  <lpage>-207</lpage>
</bibl>

<bibl id="B21">
  <title><p>Models of compensatory molecular evolution: effects of back
  mutation</p></title>
  <aug>
    <au><snm>Ichinose</snm><fnm>M.</fnm></au>
    <au><snm>Iizuka</snm><fnm>M.</fnm></au>
    <au><snm>Kusumi</snm><fnm>J.</fnm></au>
    <au><snm>Takefu</snm><fnm>M.</fnm></au>
  </aug>
  <source>Theoretical Population Biology.</source>
  <pubdate>2013</pubdate>
  <volume>323</volume>
  <fpage>1</fpage>
  <lpage>-10</lpage>
</bibl>

<bibl id="B22">
  <title><p>Stochastic tunnels in evolutionary dynamics</p></title>
  <aug>
    <au><snm>Iwasa</snm><fnm>Y.</fnm></au>
    <au><snm>Michor</snm><fnm>F.</fnm></au>
    <au><snm>Nowak</snm><fnm>M. A.</fnm></au>
  </aug>
  <source>Genetics</source>
  <pubdate>2004</pubdate>
  <volume>166</volume>
  <fpage>1571</fpage>
  <lpage>-1579</lpage>
</bibl>

<bibl id="B23">
  <title><p>Evolution of compensatory substitutions through G-U intermediate
  state in \textit{{D}rosophila} r{RNA}</p></title>
  <aug>
    <au><snm>Rousset</snm><fnm>F.</fnm></au>
    <au><snm>Pelandakis</snm><fnm>M.</fnm></au>
    <au><snm>Solignac</snm><fnm>M.</fnm></au>
  </aug>
  <source>PNAS</source>
  <pubdate>1991</pubdate>
  <volume>88</volume>
  <fpage>10032</fpage>
  <lpage>-10036</lpage>
</bibl>

<bibl id="B24">
  <title><p>The role of the effective population size in compensatory
  evolution</p></title>
  <aug>
    <au><snm>Piskol</snm><fnm>R.</fnm></au>
    <au><snm>Stephan</snm><fnm>W.</fnm></au>
  </aug>
  <source>Genome Biology and Evolution</source>
  <pubdate>2011</pubdate>
  <volume>3</volume>
  <fpage>528</fpage>
  <lpage>-538</lpage>
</bibl>

<bibl id="B25">
  <title><p>Evolutionary distances for protein-coding sequences: {M}odeling
  site-specific residue frequencies</p></title>
  <aug>
    <au><snm>Halpern</snm><fnm>A. L.</fnm></au>
    <au><snm>Bruno</snm><fnm>W. J.</fnm></au>
  </aug>
  <source>Molecular Biology and Evolution</source>
  <pubdate>1998</pubdate>
  <volume>15</volume>
  <fpage>910</fpage>
  <lpage>-917</lpage>
</bibl>

<bibl id="B26">
  <title><p>On the probability of fixation of mutant genes in a
  population</p></title>
  <aug>
    <au><snm>Kimura</snm><fnm>M.</fnm></au>
  </aug>
  <source>Genetics</source>
  <pubdate>1962</pubdate>
  <volume>47</volume>
  <fpage>713</fpage>
  <lpage>-719</lpage>
</bibl>

<bibl id="B27">
  <title><p>The {G}enetical {T}heory of {N}atural {S}election</p></title>
  <aug>
    <au><snm>Fisher</snm><fnm>R. A.</fnm></au>
  </aug>
  <publisher>Oxford: Oxford University Press</publisher>
  <pubdate>1930</pubdate>
</bibl>

<bibl id="B28">
  <title><p>Some genetic aspects of sex</p></title>
  <aug>
    <au><snm>Muller</snm><fnm>H. J.</fnm></au>
  </aug>
  <source>The American Naturalist</source>
  <pubdate>1932</pubdate>
  <volume>66</volume>
  <fpage>118</fpage>
  <lpage>-138</lpage>
</bibl>

<bibl id="B29">
  <title><p>{RNA} folding in \textit{{D}rosophila} shows a distance effect for
  compensatory fitness interactions</p></title>
  <aug>
    <au><snm>Stephan</snm><fnm>W.</fnm></au>
    <au><snm>Kirby</snm><fnm>D. A.</fnm></au>
  </aug>
  <source>Genetics</source>
  <pubdate>1993</pubdate>
  <volume>135</volume>
  <fpage>97</fpage>
  <lpage>-103</lpage>
</bibl>

<bibl id="B30">
  <title><p>Analyzing the evolution of {RNA} secondary structures in vertebrate
  introns using {K}imura's model of compensatory fitness
  interactions</p></title>
  <aug>
    <au><snm>Piskol</snm><fnm>R.</fnm></au>
    <au><snm>Stephan</snm><fnm>W.</fnm></au>
  </aug>
  <source>Molecular Biology and Evolution</source>
  <pubdate>2008</pubdate>
  <volume>25</volume>
  <fpage>2483</fpage>
  <lpage>-2492</lpage>
</bibl>

<bibl id="B31">
  <title><p>Maximum likelihood with multiparameter models of
  substitution</p></title>
  <aug>
    <au><snm>Tillier</snm><fnm>E. R. M.</fnm></au>
  </aug>
  <source>Journal of Molecular Evolution</source>
  <pubdate>1994</pubdate>
  <volume>39</volume>
  <fpage>409</fpage>
  <lpage>-417</lpage>
</bibl>

<bibl id="B32">
  <title><p>Neighbor joining and maximum likelihood with RNA sequences:
  addressing the interdependence of sites</p></title>
  <aug>
    <au><snm>Tillier</snm><fnm>E. R. M.</fnm></au>
    <au><snm>Collins</snm><fnm>R. A.</fnm></au>
  </aug>
  <source>Molecular Biology and Evolution</source>
  <pubdate>1995</pubdate>
  <volume>12</volume>
  <fpage>7</fpage>
  <lpage>-15</lpage>
</bibl>

<bibl id="B33">
  <title><p>A {D}irichlet process model for detecting positive selection in
  protein-coding {DNA} sequences</p></title>
  <aug>
    <au><snm>Huelsenbeck</snm><fnm>J. P.</fnm></au>
    <au><snm>Jain</snm><fnm>S.</fnm></au>
    <au><snm>Frost</snm><fnm>S. W. D.</fnm></au>
    <au><snm>Pond</snm><fnm>SLK</fnm></au>
  </aug>
  <source>Proceedings of the National Academy of Science, U.S.A.</source>
  <pubdate>2006</pubdate>
  <volume>103</volume>
  <fpage>6263</fpage>
  <lpage>-6268</lpage>
</bibl>

<bibl id="B34">
  <title><p>A {D}irichlet process prior for estimating lineage-specific
  substitution rates</p></title>
  <aug>
    <au><snm>Heath</snm><fnm>T. A.</fnm></au>
    <au><snm>Holder</snm><fnm>M. T.</fnm></au>
    <au><snm>Huelsenbeck</snm><fnm>J. P.</fnm></au>
  </aug>
  <source>Molecular Biology and Evolution</source>
  <pubdate>2012</pubdate>
  <volume>29</volume>
  <fpage>939</fpage>
  <lpage>-955</lpage>
</bibl>

<bibl id="B35">
  <title><p>The length of time required for a selectively neutral mutant to
  reach fixation through random frequency drift in a finite
  population</p></title>
  <aug>
    <au><snm>Kimura</snm><fnm>M.</fnm></au>
  </aug>
  <source>Genet. Res.</source>
  <pubdate>1970</pubdate>
  <volume>15</volume>
  <fpage>131</fpage>
  <lpage>-133</lpage>
</bibl>

<bibl id="B36">
  <title><p>The causes of molecular evolution</p></title>
  <aug>
    <au><snm>Gillespie</snm><fnm>J. H.</fnm></au>
  </aug>
  <publisher>New York: Oxford University Press</publisher>
  <pubdate>1991</pubdate>
</bibl>

<bibl id="B37">
  <title><p>Substitution times for mutant nucleotides</p></title>
  <aug>
    <au><snm>Watterson</snm><fnm>G. A.</fnm></au>
  </aug>
  <source>Journal of Applied Probability</source>
  <pubdate>1982</pubdate>
  <volume>19</volume>
  <fpage>59</fpage>
  <lpage>-70</lpage>
</bibl>

</refgrp>
} 


\ifthenelse{\boolean{publ}}{\end{multicols}}{}


\newpage

\section*{Figures}

\bigskip
\bigskip

\begin{figure}[htp]
\begin{center}
\includegraphics[width=110mm]{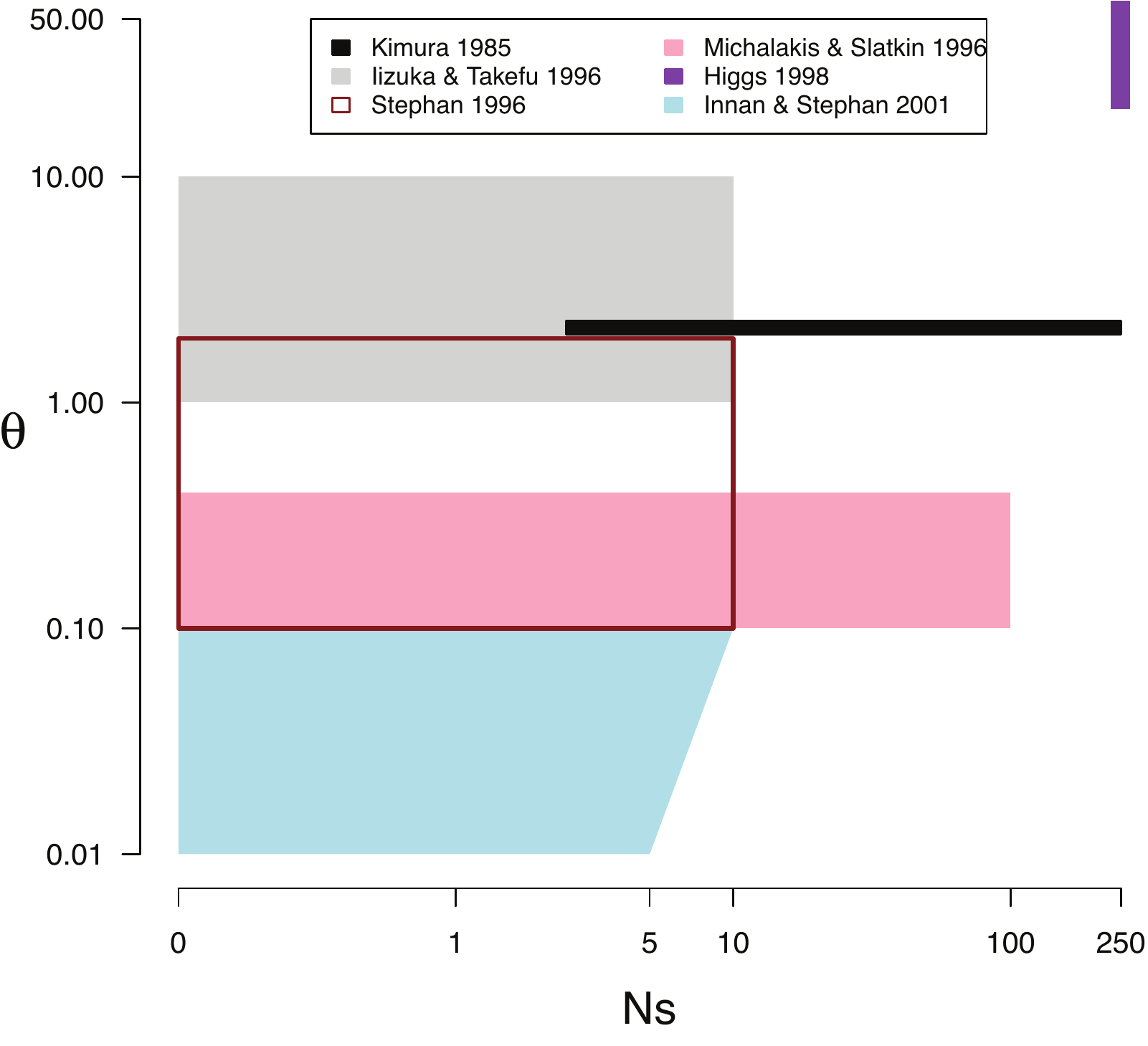}
\end{center}
\caption{The parameter ranges explored by previous work on compensatory evolution.  $\theta = 4N\mu$ where $N$ is the diploid population size, $\mu$ is the per-locus mutation rate and $s$ is the selection coefficient against deleterious intermediates.  All of these studies also explored recombination.  In black: Kimura \cite{kimura85}. In grey: Iizuka and Takefu \cite{iizuka96}. In transparent red box: Stephan \cite{stephan96}.  In pink: Michalakis and Slatkin \cite{michalakis96}.  In purple: Higgs \cite{higgs98}.  In light blue: Innan and Stephan \cite{innan01}.} 
\label{fig:paramRanges}
\end{figure}

\begin{figure}[htp]
\begin{center}
\includegraphics[width=160mm]{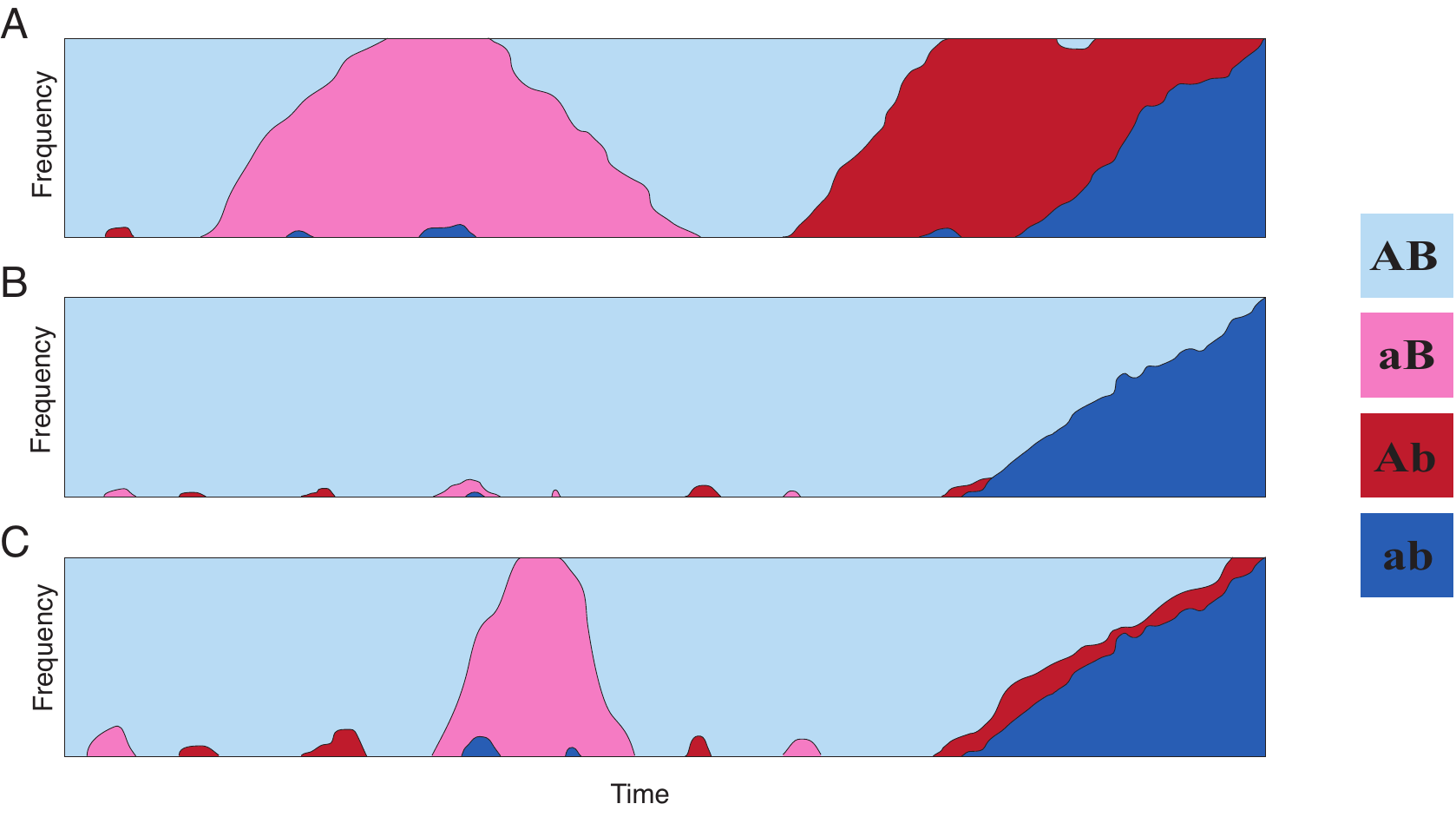}
\end{center}
\caption{Illustrations of compensatory substitution pathways.  The frequency of the two high fitness (blues) and low fitness (reds) haplotypes in a population are shown over time.  (a) Type 1. Under weak selection the deleterious intermediate can readily fix in the population.  (b) Type 2. Under strong selection the deleterious intermediates cannot rise above low frequency, and compensatory substitution must await the second mutation arising while the first is still polymorphic.  (c) Under intermediate selection both Type 1 and Type 2 events are possible.  Shown is a Type 2 event.  Although the deleterious intermediate $aB$ does become fixed, the eventual path to fixation of the double mutant $ab$ begins with $AB$ once again fixed and does not have the deleterious intermediate being fixed.}
\label{fig:pathExamples}
\end{figure}

\begin{figure}[htp]
\begin{center}
\includegraphics[width=60mm]{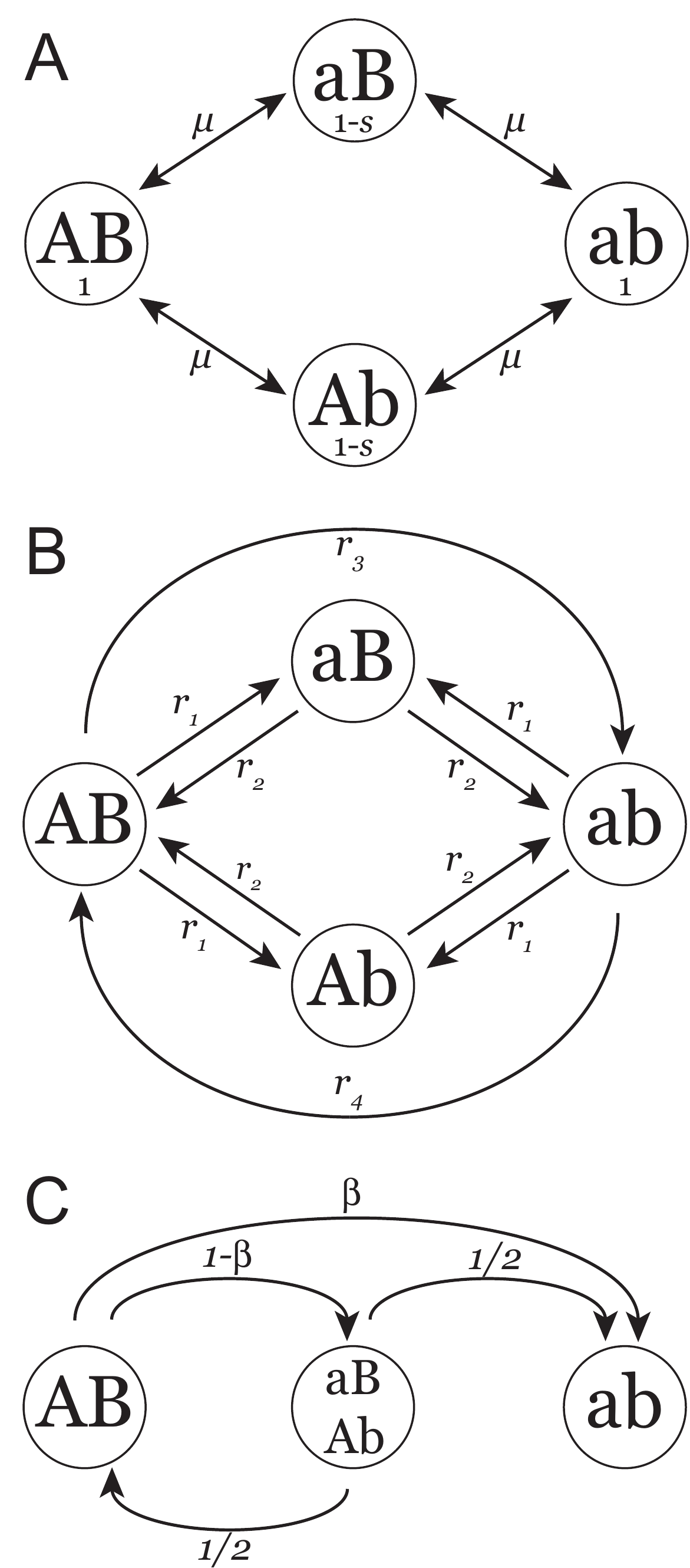}
\end{center}
\caption{Graphical model representations. (A) The mutation-selection model used.  The four haplotypes and their respective fitnesses are depicted within circles, with the mutations rates labeling the arrows.  (B) Continuous-time Markov model of the compensatory substitution process.  Rates between states are shown on arrows.  Each state represents the first fixation event for the given haplotype. (C) The discrete-time Markov chain, or jump chain, embedded within the continuous-time Markov chain. The relative probabilities of moving between fixed states are shown on arrows. Deleterious states are considered together for simplicity, and $ab$ is considered an absorbing state to match the simulation procedure.}
\label{fig:models}
\end{figure}

\begin{figure}[htp]
\begin{center}
\includegraphics[width=130mm]{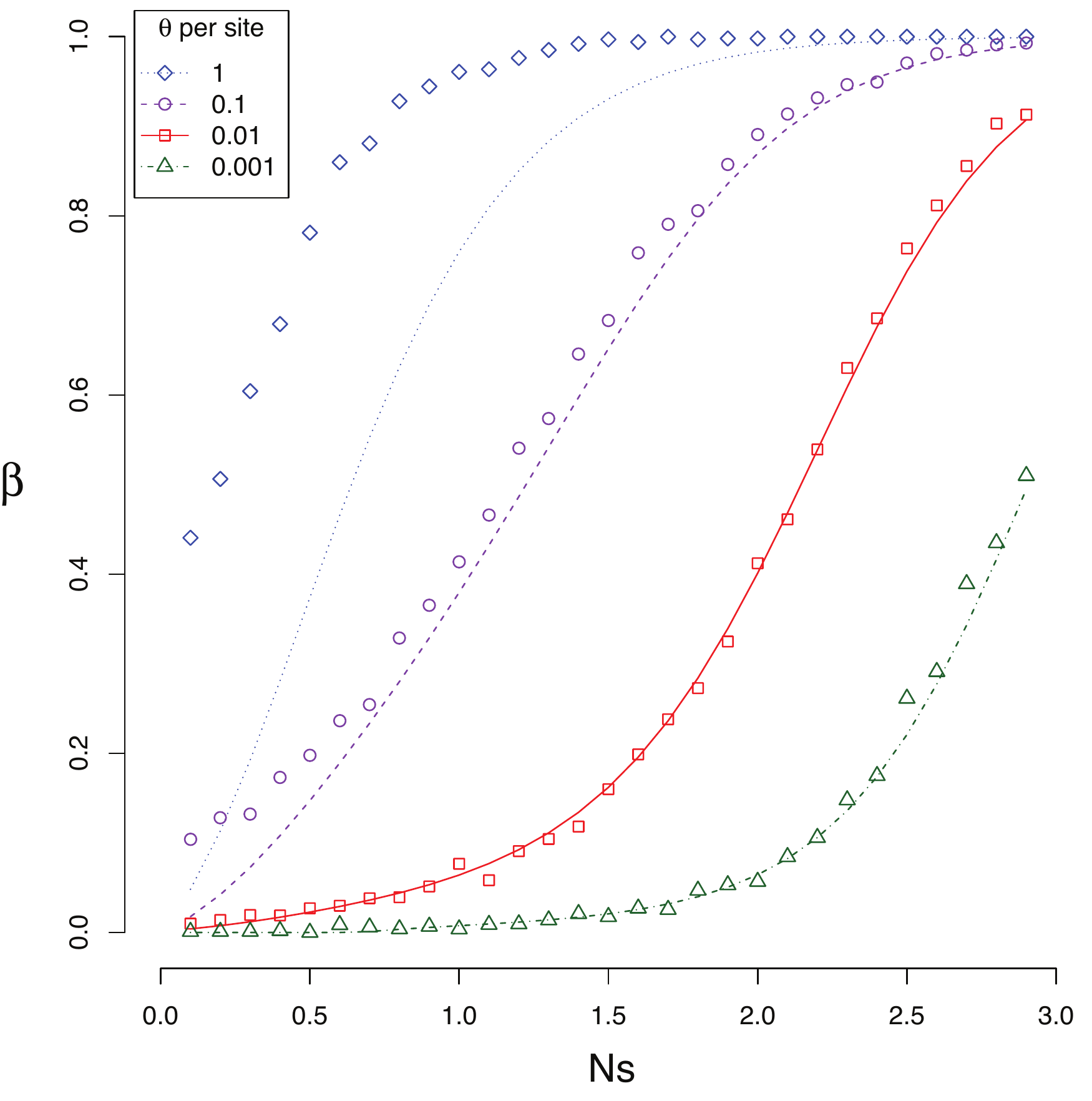}
\end{center}
\caption{The direct double substitution probability $\beta$.  The analytical predictions using Equation \ref{eq:beta} (lines) are consistent with simulations (points) for most mutation rates (colors), except when the mutation rate is very large.}
\label{fig:beta}
\end{figure}

\begin{figure}[htp]
\begin{center}
\includegraphics[width=140mm]{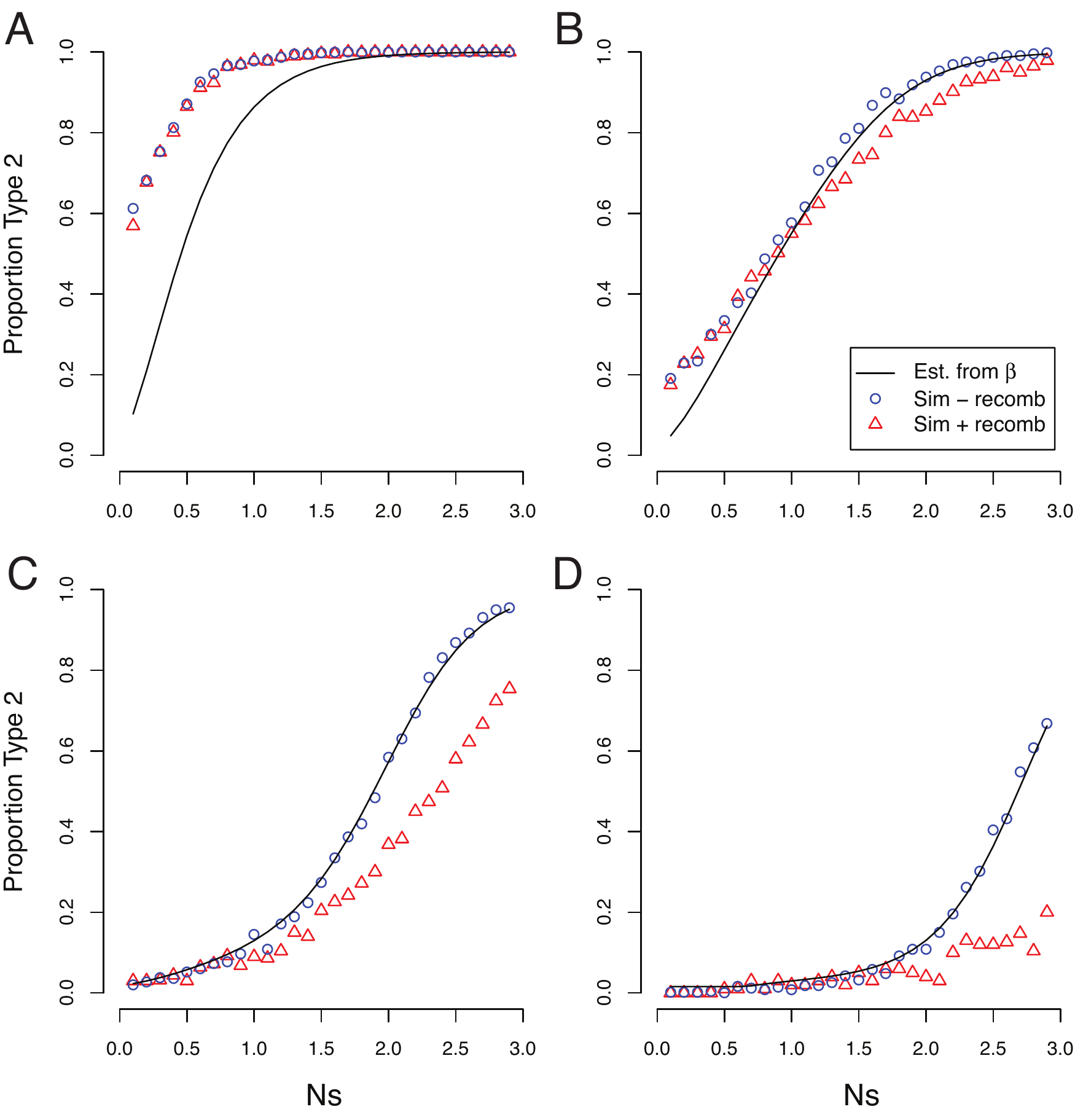}
\end{center}
\caption{Proportion of Type 2 compensatory substitutions.  The analytical prediction of the probability of Type 2 events from Equation \ref{eq:propType2} (black lines) compares favorably to simulations without recombination (blue circles).  Simulations with recombination are depicted as well (red triangles).  (A) $\theta = 1$ (B) $\theta = 0.1$ (C) $\theta = 0.01$ (D) $\theta = 0.001$}
\label{fig:propType2}
\end{figure}

\begin{figure}[htp]
\begin{center}
\includegraphics[width=110mm]{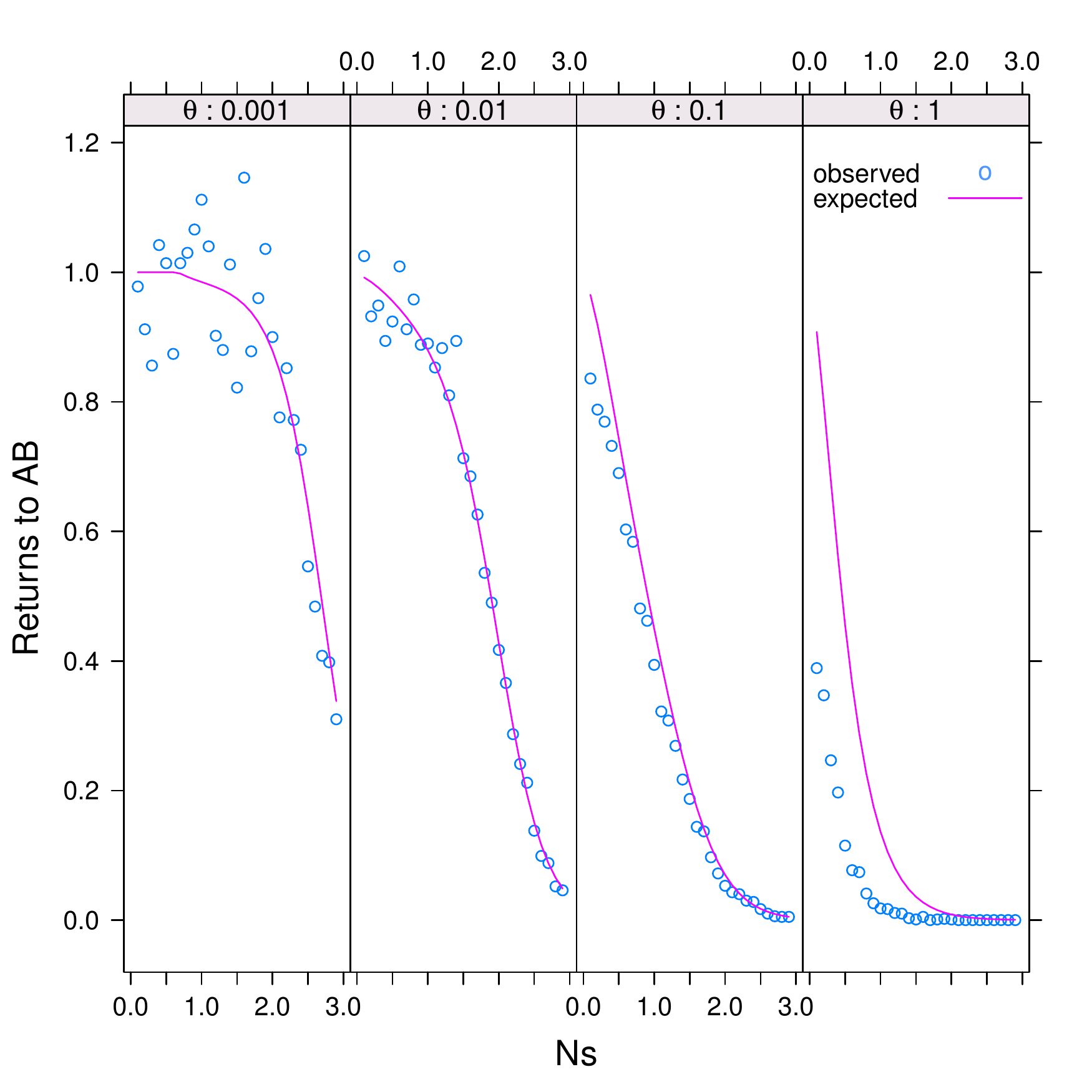}
\end{center}
\caption{Reversions to the ancestral state.  The average number of times returning to state $AB$ after having been fixed for a deleterious intermediate state are shown under four mutation rates (panels) across a variety of selection coefficients. The analytical predictions (lines) from $\beta$ of the number of recursions match well with results from simulations without recombination (points), except when the mutation rate is very large.}
\label{fig:returnsToAB}
\end{figure}

\begin{figure}[htp]
\begin{center}
\includegraphics[width=110mm]{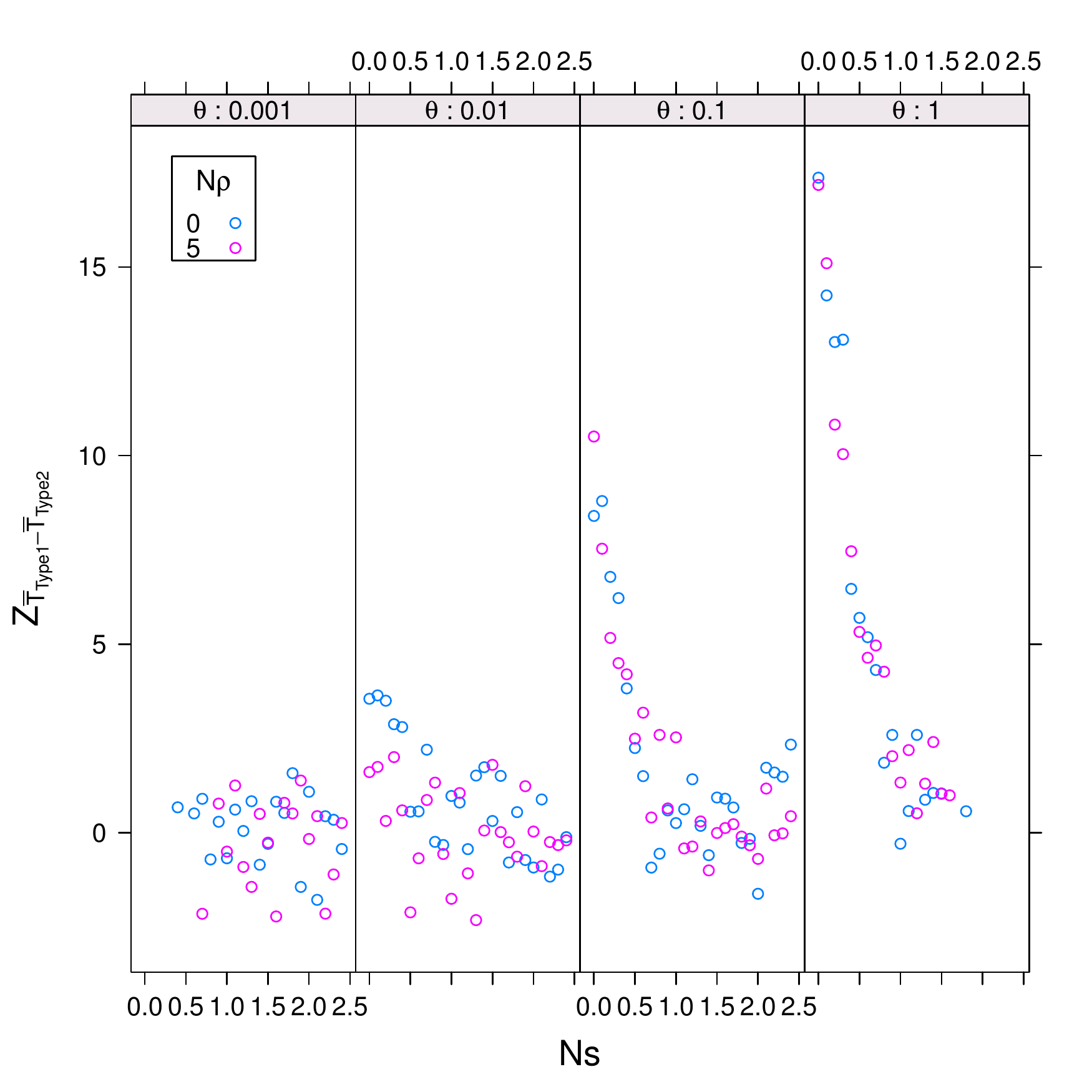}
\end{center}
\caption{The normalized time difference between Type 1 and Type 2 substitution events.  The test statistic is the difference in mean times for the two paths, normalized by the pooled standard deviation, and is shown under four mutation rates (panels) across a range of selection coefficients.  Shown are simulations in the absence (blue) and presence (pink) of recombination.}
\label{fig:lastCompSubTime}
\end{figure}

\setcounter{figure}{0}

\makeatletter
\makeatletter \renewcommand{\fnum@figure}
{\figurename~S\thefigure}
\makeatother

\begin{figure}[htp]
\begin{center}
\includegraphics[width=140mm]{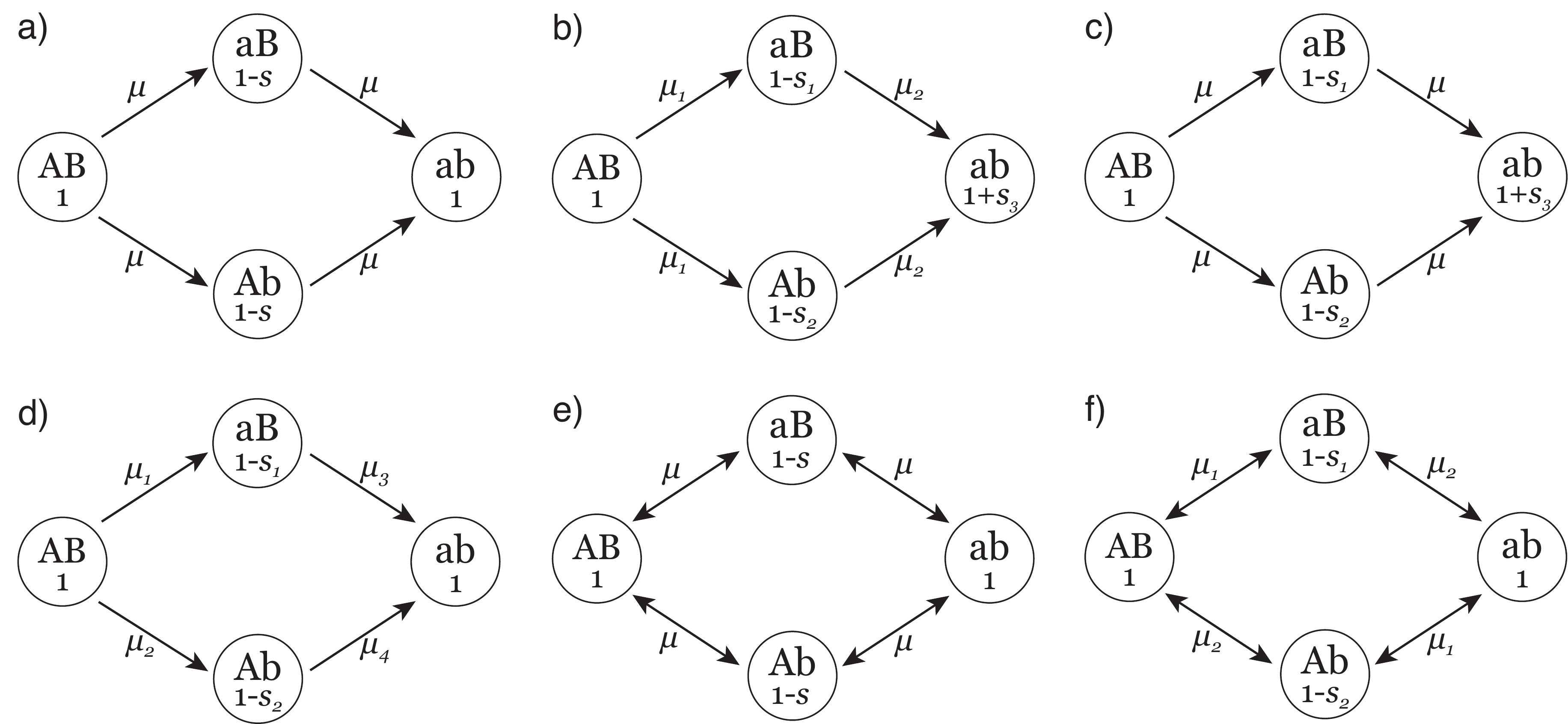} 
\end{center}
\caption{Mutation-selection models used previously for study of the compensatory substitution process in a population.
(a) Kimura (1985)
(b) Iizuka and Takefu (1996) 
(c) Michalakis and Slatkin (1996) 
(d) Stephan (1996) 
(e) Higgs (1998) 
(f) Innan and Stephan (2001)}
\label{fig:compareModels} 
\end{figure}

\begin{figure}[htp]
\begin{center}
\includegraphics[width=110mm]{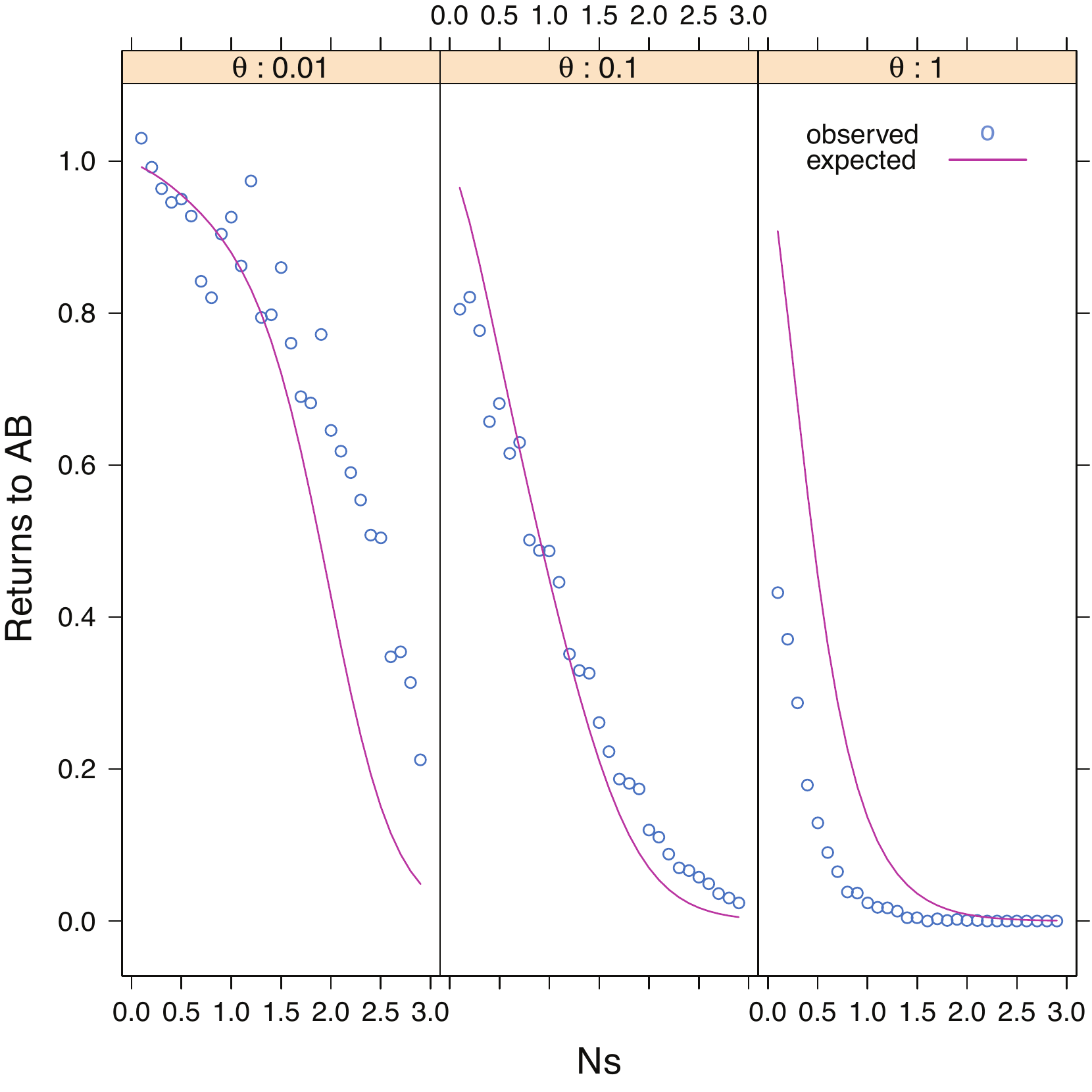}
\end{center}
\caption{Reversions to the ancestral state.  The average number of times returning to state $AB$ after having been fixed for a deleterious intermediate state are shown under four mutation rates (panels) across a variety of selection coefficients. The analytical predictions (lines) from $\beta$ of the number of recursions compared to simulations in the presence of recombination (points).  Data for $\theta=0.001$ is not shown due to low sample size.}
\end{figure}
\clearpage


\section*{Tables}


\begin{table}[htp]
\caption{Simulation Parameters}
\label{tab:WFsimParm}
\begin{center}
\begin{tabular}{lll}
Parameter 		& Symbol		&	Value \\
\hline
Mutation Rate		&	$\theta = 4N\mu$ / locus / generation &  	(0.001, 0.01, 0.1, 1.0) \\
Recombination Rate	&	$2N\rho$	 / generation	&  	(0, 5) \\
Selection Coefficient	&	$Ns$		&	(0, 0.1, 0.2, \dots , 3.0)   \\
Population Size		&	$2N$	&	200 \\
\hline
\end{tabular}
\end{center}
\end{table}


%

\end{bmcformat}
\end{document}